\documentstyle[preprint,aps]{revtex}
\begin{document}
\title
{
The orbital and charge ordering in Pr$_{1-x}$Ca$_x$MnO$_3$ ($x$=0 and 0.5)
from the {\it ab initio} calculations} \author
{V.I.Anisimov, I.S.Elfimov, M.A.Korotin} \address
{Institute of Metal Physics, Russian Academy of Sciences, 620219
Ekaterinburg GSP-170, Russia}
\author{K.Terakura}
\address{ Joint Research Center for Atomic Technology, National Institute
for Advanced Interdisciplinary Research, 1-1-4 Higashi, Tsukuba, Ibaraki
305, Japan} \date{\today}
\maketitle

\begin{abstract}
The electronic structure of the doped manganites Pr$_{1-x}$Ca$_x$MnO$_3$
was calculated by the LSDA+U method which takes into account the local
Coulomb interaction between d-electrons of transition metal ions. In
contrast to the standard local spin density approximation (LSDA) no
Jahn-Teller lattice distortions are needed to reproduce the experimentally
observed orbital ordering in the undoped PrMnO$_3$. The correct ground
state : a charge and orbital ordered antiferromagnetic insulator, was
obtained for Pr$_{1/2}$Ca$_{1/2}$MnO$_3$. The results are in good agreement
with the neutron diffraction data. 

\end{abstract}
\section{INTRODUCTION}
The doped rare earth manganites $Re_{1-x}A_x$MnO$_3$ ($Re$ is a rare earth
such as La and $A$ is a divalent element such as Sr or Ca) due to their
peculiar correlation between magnetism and conductivity have been
extensively studied during 1950s and 60s \cite{classic}. The most throughly
investigated was the La$_{1-x}$Sr$_x$MnO$_3$ system. Undoped ($x$=0),
LaMnO$_3$ is an antiferromagnetic insulator. Upon doping with Sr, this
perovskite oxide becomes a ferromagnetic metal; the connection between
metallicity and ferromagnetism was well explained by the double exchange
hopping mechanism \cite{doublex}. The discovery of colossal
magnetoresistance phenomena in samples with Sr dopant densities in the
0.2$\leq x \leq$0.4 regime \cite{gmr} brought a revival of the interest to
these sytems.

The Mn$^{+3}$ ion in the hole-undoped compound LaMnO$_3$ has the high-spin
$d^4$ electron configuration $t_{2g\uparrow}^3 e_{g\uparrow}^1$. The
$t_{2g}$ orbitals hybridize with O2p orbitals much more weakly than the
$e_g$ orbitals and
can be regarded as forming the localized spin ($S$=3/2). In contrast to
that, $e_g$ orbitals, which have lobes directed to the neighboring oxygen
atoms, hybridize strongly with O2p
 producing in the result rather broad bands. The
strong exchange interaction with $t_{2g\uparrow}$ subshell leads to the
splitting of the $e_g$ band into unoccupied $e_{g\downarrow}$ and
half-occupied $e_{g\uparrow}$ subbands. The half filled $e_{g\uparrow}$
subband is a typical example of the Jahn-Teller system, and, indeed,
LaMnO$_3$ has an orthorombic crystal structure \cite{crstr} with distorted
(elongated) MnO$_6$ octahedra. This cooperative Jahn-Teller effect 
is usually considered responsible for the
opening of the gap in the half-filled $e_{g\uparrow}$ band and an
insulator ground state of LaMnO$_3$. 

The orthorombic crystal structure of LaMnO$_3$ ($Pbnm$ space group) can be
described as a perovskite with two types of distortion from a cubic
structure: 1)Tilting (rotation) of the MnO$_6$ octahedra, so that Mn-O-Mn
angles become less than 180$^\circ$, and 2)Jahn-Teller distortion of the
octahedra, with one long Mn-O bond and two short bonds. With doping by Sr
the orhtorombicity (the deviation of the values of the lattice parameters
$b$ and $c/\sqrt{2}$ from $a$) becomes smaller and for $x\geq 0.175$ the
stable crystal structure at room temperature becomes rombohedral, where
only tilting of oxygen octahedra is present but all three Mn-O bonds are
equal. Around the same value of the doping $x$, La$_{1-x}$Sr$_x$MnO$_3$
becomes metallic. 

For Sr doping with 0.2$\leq x \leq$0.4, La$_{1-x}$Sr$_x$MnO$_3$ is a
ferromagnetic metal at low temperatures. However, as the temperature
increases and approaches the Curie temperature $T_c$ a sudden increase of
the resistivity is observed \cite{gmr}. As the temperature dependence of
the resistivity $\rho(T)$ has typically semiconducting behavior ( $d\rho /
dT < 0$) above T$_c$, this increase in resistivity at around $T_c$ is
usually desribed as a metal-insulator transition. The colossal
magnetoresistance effect happens when the temperature is close to $T_c$:
the external magnetic field leads to suppression of the resistivity 
increase or to
shifting the metal-insulator transition to higher temperatures resulting in
a very large negative value of $[\rho(H)-\rho(0)]/\rho(0)$. 

The crucial point in the understanding of the effect of the colossal
magnetoresistance is the nature of the metal-insulator transition with the
temperature variation across $T_c$. It was shown \cite{millis1} that
the double exchange mechanism alone is not enough to explain such
transition.  Millis {\it et al.} \cite{millis2} suggested local Jahn-Teller
distortions (Jahn-Teller polarons) as a main mechanism causing
localization. Their idea is that a $E_{JT}/W$ ratio ($E_{JT}$ is
self-trapping Jahn-Teller polaron energy and $W$ is effective band width)
is close to the critical value in La$_{1-x}$Sr$_x$MnO$_3$ and as $W$
decreases with destroying the ferromagnetic order above $T_c$ (the
effective hopping takes the maximum value for a parallel spin alignment),
this ratio becomes larger than the critical value to result in polaron
localization.
Varma \cite{varma} argues that LaMnO$_3$ is not an insulator due to the
Jahn-Teller distortion but is a Mott insulator and the Jahn-Teller
distortion occurs parasitically. He explains localisation of the holes in
the paramagnetic phase by spin polaron mechanism. 

By replacing La with other trivalent ions with smaller ionic radius, the
Mn-O-Mn bond angle becomes smaller and the $e_g$ band width $W$ is reduced
to enhance the tendency to the  carrier localization and lattice distortion. 
Variety of dramatic phenomena have been observed in (Pr, Nd, Sm)(Sr,
Ca)MnO$_3$ systems.  As one of such examples, we study
Pr$_{1/2}$Ca$_{1/2}$MnO$_3$ in the present paper.  This system has a very
peculiar phase diagram \cite{jirak,prcamno3}.  At low temperature, it is an
antiferromagenetic insulator with a charge ordering of Mn$^{3+}$ and
Mn$^{4+}$ accompanied by an orbital ordering: the orbital ordering and spin
ordering is of CE-type.  Between $T_{\rm N}$ ($\simeq 180$ K) and $T_{\rm
co}$ ($\simeq 240$ K), there is no long-range magnetic ordering but the
system remains to be insulating because of the persisitence of the charge
ordering.  Above $T_{\rm co}$, it is a paramagnetic insulator.

The local spin density approximation (LSDA)\cite{lda} 
was used by several
groups for theoretical investigation of
the electronic structure of La$_{1-x}$Sr$_x$MnO$_3$
\cite{hamada,pickett,satpathi}. Those calculations have confirmed the
importance of the Jahn-Teller distortion for correct description by the
LSDA of the insulating antiferromagnetic ground state of the undoped
LaMnO$_3$, because the calculations with an undistorted 
cubic perovskite crystal
structure produce a half-filled metallic $e_{g\uparrow}$ band. However in
order to address the charge and orbital order observed in doped manganites
it is necessary to go beyond LSDA by including the intra d-shell Coulomb
interaction (LSDA+U method \cite{ldau,ANISOL,lichtan}). Recently such an
approach was used for treating the charge ordering in magnetite
Fe$_3$O$_4$, where LSDA gives a metallic uniform solution. The LSDA+U
calculation gave a charge ordered insulator as the ground state of this
compound \cite{fe3o4}.
In this paper we describe the results of the electronic structure
calculation for Pr$_{1-x}$Ca$_x$MnO$_3$ for $x=0$ and 0.5 in the LSDA+U
approximation.

\section{COMPUTATIONAL METHOD}
The problem of the charge ordering cannot be treated properly by the
standard LSDA because of the non-vanishing self-interaction. In
contrast to the
Hartree-Fock approximation, where the self-interaction is explicitly
excluded for every orbital, in the
LSDA it is canceled to a good approximation in the total energy, but rather
poorly in the local one-electron potentials, 
which are orbital-independent.
The spurious self-interaction present in the LSDA leads to an increase in the
Coulomb interaction when the
distribution of the electron charge deviates from the uniform one. This
effect can be illustrated in the following way. If we consider the system
having less than one electron per site in the partially filled electronic
subshell (for example $e_g$-electrons in La$_{1-x}$Sr$_x$MnO$_3$) then the
potential which this electron feels at some site must not depend on the
orbital occupancy of this particular site, as the electron does not
interact with itself. However, as the LSDA
potential is a functional of the electron density only, increasing the
electron density at one site and decreasing it at another one, with a
formation of the charge ordering, will lead to an increase of the potential
at the first site and a
decrease at the second one. As a
result,
in the self-consistency loops, the charge distribution will return to the
uniform density.
In order to cure this deficiency, it is necessary to remove the
self-interaction.
The LSDA+U method \cite{ldau,lichtan} is one of the practical methods to
solve the self-interaction problem.

Let us introduce the d-orbitals occupation matrix defined by: 

\begin{equation}
\label{Occ}n_{mm^{\prime }}^\sigma =-\frac 1\pi \int^{E_F} {\rm
{Im}}G_{inlm,inlm^{\prime }}^\sigma (E)dE, \end{equation}
where $G_{inlm,inlm^{^{\prime }}}^\sigma (E)=\langle inlm\sigma \mid (E-
\widehat{H})^{-1}\mid inlm^{^{\prime }}\sigma \rangle $ are the elements of
the Green function matrix ($i$: index of atom site, $nl$: main and orbital
quantum numbers (for example 3$d$), $m$: index of a particular $d$-orbital
and $\sigma$: spin projection index).
In terms of the elements of this occupation matrix $\{n^\sigma \}$, the LSDA+U 
functional \cite{lichtan} is defined as follows:

\begin{equation}
\label{U1}E^{LSDA+U}[\rho ^\sigma ({\bf r}),\{n^\sigma \}]=E^{LSDA}[\rho
^\sigma ({\bf r)}]+E^U[\{n^\sigma \}]-E_{dc}[\{n^\sigma \}] \end{equation}
Where $\rho ^\sigma ({\bf r})$ is the charge density for spin-$\sigma $
electrons and $E^{LSDA}[\rho ^\sigma ({\bf r})]$ is the standard LSDA
(Local Spin-Density Approximation) functional.  $E^U[\{n_{\sigma}\}]$ is
the Hartree-Fock type electron-electron interaction energy given by
\begin{equation}
\label{upart}
\begin{array}{c}
E^U[\{n\}]=\frac 12\sum_{\{m\},\sigma }\{\langle m,m^{\prime \prime }\mid
V_{ee}\mid m^{\prime },m^{\prime \prime \prime }\rangle n_{mm^{\prime
}}^\sigma n_{m^{\prime \prime }m^{\prime \prime \prime }}^{-\sigma } \\ \\
-(\langle m,m^{\prime \prime }\mid V_{ee}\mid m^{\prime },m^{\prime \prime
\prime }\rangle -\langle m,m^{\prime \prime }\mid V_{ee}\mid m^{\prime
\prime \prime },m^{\prime }\rangle )n_{mm^{\prime }}^\sigma n_{m^{\prime
\prime }m^{\prime \prime \prime }}^\sigma \}, \end{array}
\end{equation}
where $V_{ee}$ is the screened Coulomb interaction among the $nl$
electrons. Finally, the last term in Eq.(\ref{U1}) corrects for double
counting and is given by

\begin{equation}
\label{U3}E_{dc}[\{n^\sigma \}]=\frac 12UN(N-1)-\frac 12J[N^{\uparrow
}(N^{\uparrow }-1)+N^{\downarrow }(N^{\downarrow }-1)], \end{equation}
were $N^\sigma =Tr(n_{mm^{\prime }}^\sigma )$ and $N=N^{\uparrow
}+N^{\downarrow }.$ $U$ and $J$ are screened Coulomb and exchange
parameters \cite{superlsda,anigun}.

In addition to the usual LSDA potential,
the variation of the functional (\ref{U1}) gives an effective single-particle
potential to be used in the 
single-particle Hamiltonian :

\begin{equation}
\label{hamilt}\widehat{H}=\widehat{H}_{LSDA}+\sum_{mm^{\prime }}\mid
inlm\sigma \rangle V_{mm^{\prime }}^\sigma \langle inlm^{\prime }\sigma
\mid \end{equation}
\begin{equation}
\label{Pot}
\begin{array}{c}
V_{mm^{\prime }}^\sigma =\sum_{m^{\prime \prime} m^{\prime \prime
\prime}}\{\langle m,m^{\prime \prime }\mid V_{ee}\mid m^{\prime },m^{\prime
\prime \prime }\rangle n_{m^{\prime \prime }m^{\prime \prime \prime
}}^{-\sigma } \\ \\
+(\langle m,m^{\prime \prime }\mid V_{ee}\mid m^{\prime },m^{\prime \prime
\prime }\rangle -\langle m,m^{\prime \prime }\mid V_{ee}\mid m^{\prime
\prime \prime },m^{\prime }\rangle )n_{m^{\prime \prime }m^{\prime \prime
\prime }}^\sigma \} \\
\\
-U(N-\frac 12)+J(N^{\sigma}-\frac 12).
\end{array}
\end{equation}

The matrix elements of the screened Coulomb interaction $V_{ee}$ can be
expressed in terms of complex spherical harmonics and effective Slater
integrals $F^k$\cite{JUDD} as
\begin{equation}
\label{slater}\langle m,m^{\prime \prime }\mid V_{ee}\mid m^{\prime
},m^{\prime \prime \prime }\rangle =\sum_ka_k(m,m^{\prime },m^{\prime
\prime },m^{\prime \prime \prime })F^k,
\end{equation}
where $0\leq k\leq 2l$ and

$$
a_k(m,m^{\prime },m^{\prime \prime },m^{\prime \prime \prime })=\frac{4\pi
}{ 2k+1}\sum_{q=-k}^k\langle lm\mid Y_{kq}\mid lm^{\prime }\rangle \langle
lm^{\prime \prime }\mid Y_{kq}^{*}\mid lm^{\prime \prime \prime }\rangle $$

For d-electrons one needs $F^0,F^2$ and $F^4$ and these can be linked to
the Coulomb- and Stoner parameters $U$ and $J$ obtained by the
LSDA-supercell procedures
\cite{superlsda,anigun}
via $U=F^0$ and $J=(F^2+F^4)/14$ with the ratio $F^2/F^4$ being to a good
accuracy a constant $\sim 0.625$ for the 3d elements \cite{ANISOL,deGroot}.

The LSDA+U approximation was realized in the framework of the linear
muffin-tin orbital method in the atomic sphere approximation
(LMTO-ASA)\cite{lmto}. The values of $U$=7.9 eV and $J$=0.9 eV were used as
obtained from the constrained LSDA supercell calculations. 

\section{RESULTS AND DISCUSSION}

As a first step
the electronic structure of the undoped PrMnO$_3$ was calculated. This
compound has practically the same properties as LaMnO$_3$. The only
difference is that due to the smaller ionic radius of Pr ion compared with
that of La ion, the tilting of the oxygen octahedra in orthorombic $Pbnm$
crystal structure is stronger, resulting in smaller effective $e_g-e_g$
hopping between Mn atoms and narrower band width.

The crystal structure of the doped
Pr$_{1-x}$Ca$_x$MnO$_3$ does not show transition to the rombohedral
symmetry, as it is the case for
La$_{1-x}$Sr$_x$MnO$_3$, preserving the orthorombic $Pbnm$ space
group\cite{jirak}. However the values of the lattice parameters $a$,$b$ and
$c/\sqrt{2}$ for $x$=0.5 are so close that it can be called a pseudocubic
structure. In this structure only the tilting of the oxygen octahedra is
present, but octahedra themselves are not Jahn-Teller distorted as
in the undoped crystal structure.
In order to study effects of the orbital
polarization purely due to the intrashell d-d Coulomb interaction without
any influence of the Jahn-Teller lattice distortion the calculation for the
undoped PrMnO$_3$ was performed with the crystal structure parameters
corresponding to Pr$_{1/2}$Ca$_{1/2}$MnO$_3$.

Kugel and Khomskii \cite{KKh} demonstrated for a wide range of Jahn-Teller
magnetic compounds that the exchange mechanism (caused by an interplay
between on-site Coulomb interaction $U$ and intersite hopping $t$)
by itself, without taking
Janh-Teller distortions into consideration, is able to give a correct
picture of the spin and orbital ordering. In this case the change in the
lattice structure (the structural transition) is a secondary effect. One
may say that the system shows the "Jahn-Teller ordering" without the
"Jahn-Teller interaction". Actually an exchange mechanism acts jointly with
the  Jahn-Teller and the quadrupole mechanisms. It is quite difficult to
determine just which mechanism will be predominant in each specific case
due to complexity of the real compound. One of the possible ways to answer
this question in the frame of {\it ab-initio} band structure calculations
is
to use a crystal structure excluding
distortions caused by the Jahn-Teller interaction. 

Such kind of approach was used for the cooperative Jahn-Teller system
KCuF$_3$ \cite{lichtan}. It also has the perovskite crystal structure, but
there is no tilting of the CuF$_6$ octahedra and all Cu-F-Cu angles are
equal to 180$^\circ$. Divalent copper ion in this compound has a $d^9$
configuration ($t_{2g\uparrow}^3 t_{2g\downarrow}^3 e_{g\uparrow}^2
e_{g\downarrow}^1$) known to be a Jahn-Teller ion, and, indeed, the
Jahn-Teller distortion is observed
in the form of the elongated CuF$_6$ octahedra with long axes alternatively
along $a$ and $b$ axes in the $ab$-plane. Kugel and Khomskii \cite{KKh}
showed that a simple Hubbard model type calculation in a mean-field
approximation without taking into account the electron-phonon interaction
could predict the orbital ordering (alternating $3x^2-r^2$ and $3y^2-r^2$
Cu3d-orbitals for electrons or $x^2-z^2$ and $y^2-z^2$ orbitals for holes)
corresponding to the symmetry of the experimentally observed lattice
distortions.

The LSDA+U functional (eq.\ref{U1}) has a definite important improvement
upon the LSDA: explicit cancellation of the self-interaction and the proper
description of the lowering of Coulomb interaction energy with development
of the orbital polarization. The functional (eq.\ref{U1}) is also
rotationally invariant (the total energy does not change with the unitary
transformation of the d-orbital basis set $|m>$), meaning that if the local
coordinate system (and, correspodingly d-orbital basis set
$|m>$) was chosen inappropriately for describing the resulting orbital
polarisation, the unitary transformation which diagonalizes  the
non-diagonal occupation matrix $n_{mm^{\prime }}^\sigma$ (eq.\ref{Occ})
will give a proper orbital basis set.

The LSDA+U calculation for KCuF$_3$ \cite{lichtan} in {\it cubic}
perovskite crystal structure gave results fully confirming Kugel-Khomskii's
prediction - alternating $3x^2-r^2$ and $3y^2-r^2$ orbital order. Moreover,
it was not just the lowest total energy configuration among many possible
stationary configurations as
in the model calculation \cite{KKh}, but the {\it only one stable
solution}. The lattice relaxation to minimize the total energy is in good
agreement with experimental data.   Note that the standard LSDA did not
show the lattice instability with respect to the Jahn-Teller distortion in
KCuF$_3$. We would like to emphasize that the orbital polarization caused
by the d-d Coulomb interactions practically did not change with the lattice
distortion \cite{sawada}. 

The use of the crystal structure parameters corresponding to
Pr$_{1/2}$Ca$_{1/2}$MnO$_3$ (with undistorted oxygen octahedra)
for PrMnO$_3$ may correspond to the use of
the cubic perovskite crystal structure for KCuF$_3$. However, for
PrMnO$_3$, we have to take into account the tilting of MnO$_6$ octahedra
(present in Pr$_{1/2}$Ca$_{1/2}$MnO$_3$ structure)                        
,
which is essential for the proper description of the reduction in the
effective $e_g-e_g$ hopping due to the bending of the Mn-O-Mn bond. The
LSDA+U calculation with the experimental A-type
antiferromagnetic spin alignment resulted in a non-diagonal occupation
matrix for spin density (for the $e_{g}$ subspace 
($3z^2-r^2$ and $x^2-y^2$ orbitals) of one particular Mn atom):
\begin{equation}
n_{mm'}^\uparrow-n_{mm'}^\downarrow
=\left(\begin{array}{cc}0.31&0.22\\0.22&0.58\end{array} \right)
\end{equation}
The diagonalization of this matrix
gives two new $e_{g\uparrow}$
orbitals: $\phi_1=3y^2-r^2$ with occupancy 0.71 and $\phi_2=z^2-x^2$ with
occupancy 0.18. For the second type of Mn atom $\phi_1$ and $\phi_2$
orbitals can be obtained by transposition of $x$ and $y$. 

The resulting orbital order can be presented graphically by plotting the
angle distribution of the $e_g$-electron spin density: \begin{equation}
\label{f1}
\rho(\theta,\phi)=\sum_{mm'}(n_{mm'}^\uparrow
-n_{mm'}^\downarrow)Y_m(\theta,\phi)Y_{m'}(\theta,\phi) \end{equation}
As a Mn$^{+3}$ ion has formally only one electron in partially filled
$e_{g\uparrow}$ subshell, the $e_g$-spin density must correspond to the
density of this electron (Fig.1). 

The plot of Fig.1 and the form of the $\phi_1$ and $\phi_2$ orbitals
diagonalizing the spin density
occupation matrix show that the orbital order obtained in
our LSDA+U calculations for PrMnO$_3$ is of the same symmetry as the one in
KCuF$_3$. The "parasitical" lattice distortion, which one could expect from
such orbital ordering: elongation of MnO$_6$ octahedra in the direction of
the lobes of $\phi_1$ orbitals, will reproduce the experimentally observed
Jahn-Teller lattice distortion. 

The orbital polarization is far from being 100\%, the difference in the
occupancy of $\phi_1$ and $\phi_2$ orbitals is only 0.53. The reason for
this is the strong hybridization of $e_{g\uparrow}$-orbitals with oxygen
2p-orbitals. In Fig.2, the total and partial densities of states (DOS) for
PrMnO$_3$ are shown (for $e_g$-electrons DOS projected on $\phi_1$ and
$\phi_2$ orbitals are also presented). 

The electronic structure obtained in our LSDA+U calculation is
semiconducting with a band gap of 0.5 eV.   Note that the standard LSDA
needs the Jahn-Teller distorted crystal structure to reproduce the
non-metallic ground state. The $e_{g\uparrow}$-band is split by this gap
into two subbands: the occupied one with the predominantly $\phi_1$
character, and the empty one with the $\phi_2$ character. However there is
significant admixture of $\phi_2$ in the occupied subband and $\phi_1$ in
the empty one. 

After checking that our calculation scheme gives a proper orbital order in
the undoped PrMnO$_3$, we applyed it to Pr$_{1/2}$Ca$_{1/2}$MnO$_3$. In
order to treat the experimentally observed CE-type of antiferromagnetism it
was necessary to quadruple the $Pbnm$-type unit cell and the supercell had
16 formula units. The distribution of the Pr and Ca atoms in the lattice
was chosen according to a model where all Pr ions have only Ca as their
nearst neighbores in the lanthanide sublattice. We did not impose any
symmetry restrictions in our calculation, the integration in the  
reciprocal space being performed for the whole Brillouin zone,
and the self-consistency iteration was started from a uniform distribution
of $e_g$-electrons over both $e_g$-orbitals of all Mn atoms.

The result of the self-consistent calculation was a charge and orbital
ordered insulator. The $e_g$-electron spin-density-plot calculated
by eq.\ref{f1} is presented in Fig.3. This is in a striking
agreement with the orbital order derived in \cite{jirak} from the
neutron diffraction mesuarements (Fig.4). The unexpected result
is that the total number of d-electrons 
at all types of Mn sites are nearly equal (4.99 and 5.01), so that formally
Mn$^{4+}$ ions (in the Fig.3 the ones with symmetric in-plane density
distribution) and Mn$^{3+}$ ions (the ones with the
density strongly anisotropic) have nearly the same number of
3d-electrons. However the difference in the magnetic moment values
is more pronounced : 3.34$\mu_B$ for Mn$^{4+}$ and 3.44 $\mu_B$ for
Mn$^{3+}$. 

\section{CONCLUSION}
The results of our LSDA+U calculations for Pr$_{1-x}$Ca$_x$MnO$_3$
demonstrate that taking into account 
the Coulomb interaction inside the d-shell of Mn 
together with the intersite hopping is enough to
reproduce not only the orbital order present in the undoped manganites,
but also the localization and ordering of the holes in the doped
materials. Naturally the lattice relaxation (Jahn-Teller and
polaron lattice distortions) will significantly renormalize
the total energies, gap values, charge disproportion 
and magnetic moments. However
it seems that the symmetry of the ordered state can be obtained
without explicitly including the electron-lattice interactions.  This is
in clear contrast to
the standard LSDA where energy band splitting and
orbital polarization appear only as results of the Jahn-Teller crystal
structure distortion.

\section{ACKNOWLEDGMENTS}
V.I.A. and M.A.K. acknowledge the NAIR Guest Research
Program. The present work is supported partly by Russian Foundation
for Basic Research (RFBR grant 96-02-16167), New Energy and Industrial
Development Organization in Japan and also by a Grant-in-Aid for Scientific
Research in the priority area "Anomalous Metallic States Near the Mott
Transition" from the Ministry of Education, Science and Culture of Japan.

\subsection*{Figure captions}
Fig. 1. The calculated angle distribution of the $e_{g\uparrow}$-electron
spin density in PrMnO$_3$.

Fig. 2. The total and partial densities of states (DOS) from the result
of LSDA+U calculation of PrMnO$_3$. a)Total DOS per unit cell (4 formula
units); b)Partial $e_g$ DOS of Mn. Solid lines are for the
$\phi_1$-projected DOS,
and dashed lines for the $\phi_2$-projected DOS with $\phi_1$ and $\phi_2$
-orbitals
diagonalizing the occupation matrix.; c) and d) Partial DOS for two types of 
oxygen orbitals.

Fig. 3. The calculated angle distribution of the $e_{g\uparrow}$-electron
spin density in Pr$_{1/2}$Ca$_{1/2}$MnO$_3$.

Fig. 4. The scheme of the spin, charge and orbital order 
in Pr$_{1/2}$Ca$_{1/2}$MnO$_3$ deduced from
the neutron difraction data \cite{jirak}.  The open circles with the lobe
of the $e_g$ electron density distribution denote Mn$^{3+}$ and the filled
circles Mn$^{4+}$.  The arrows denote the magnetic moments.
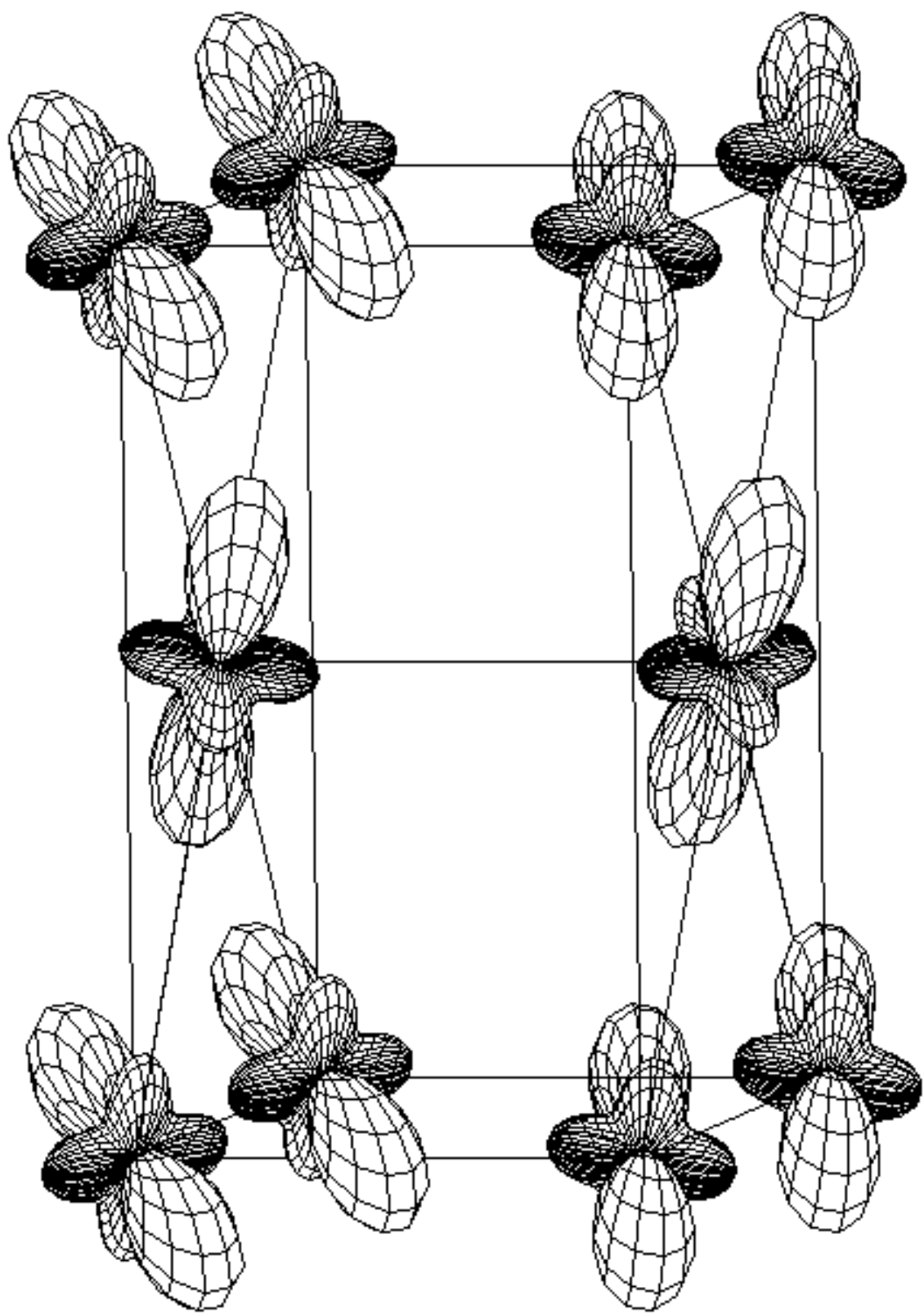
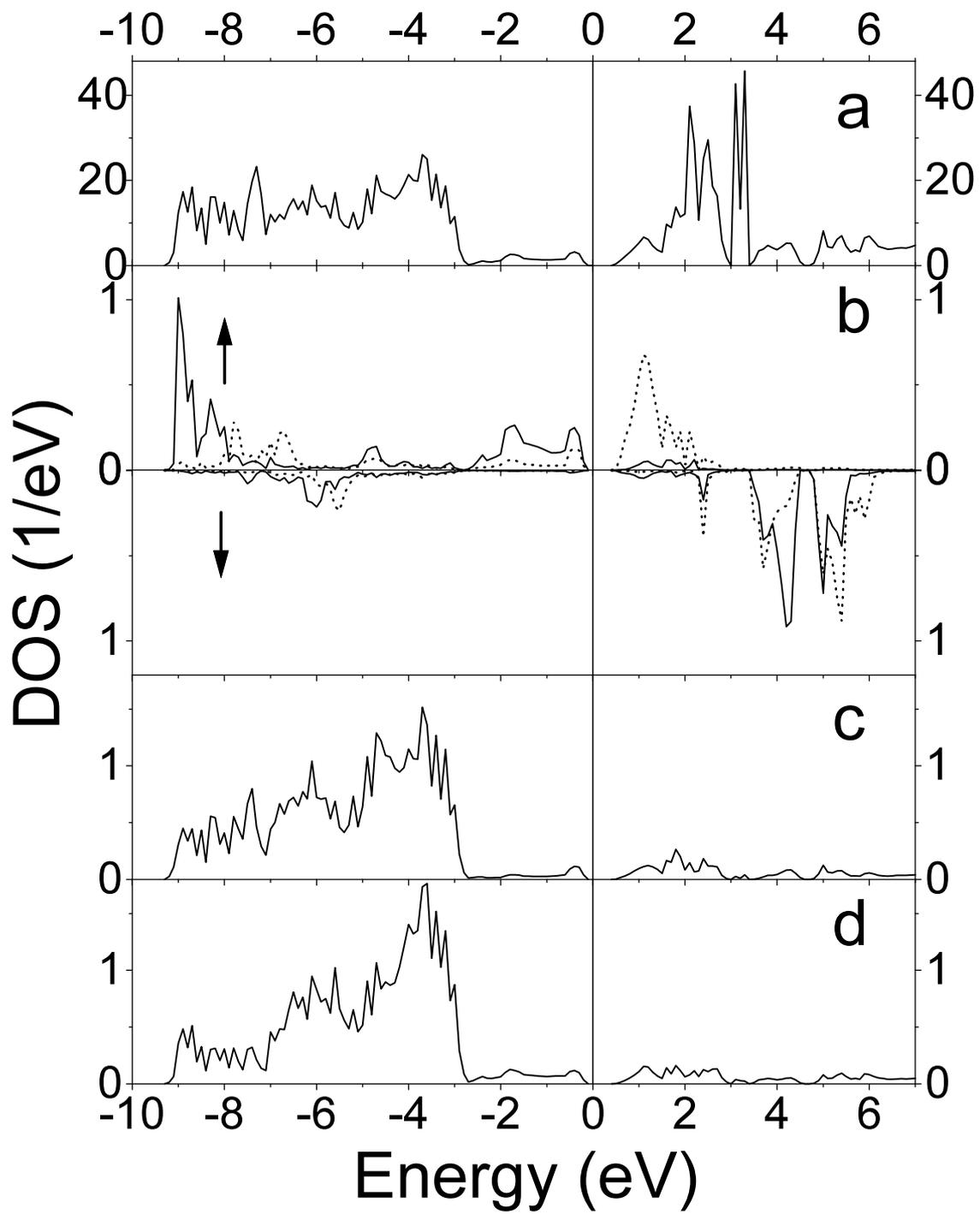
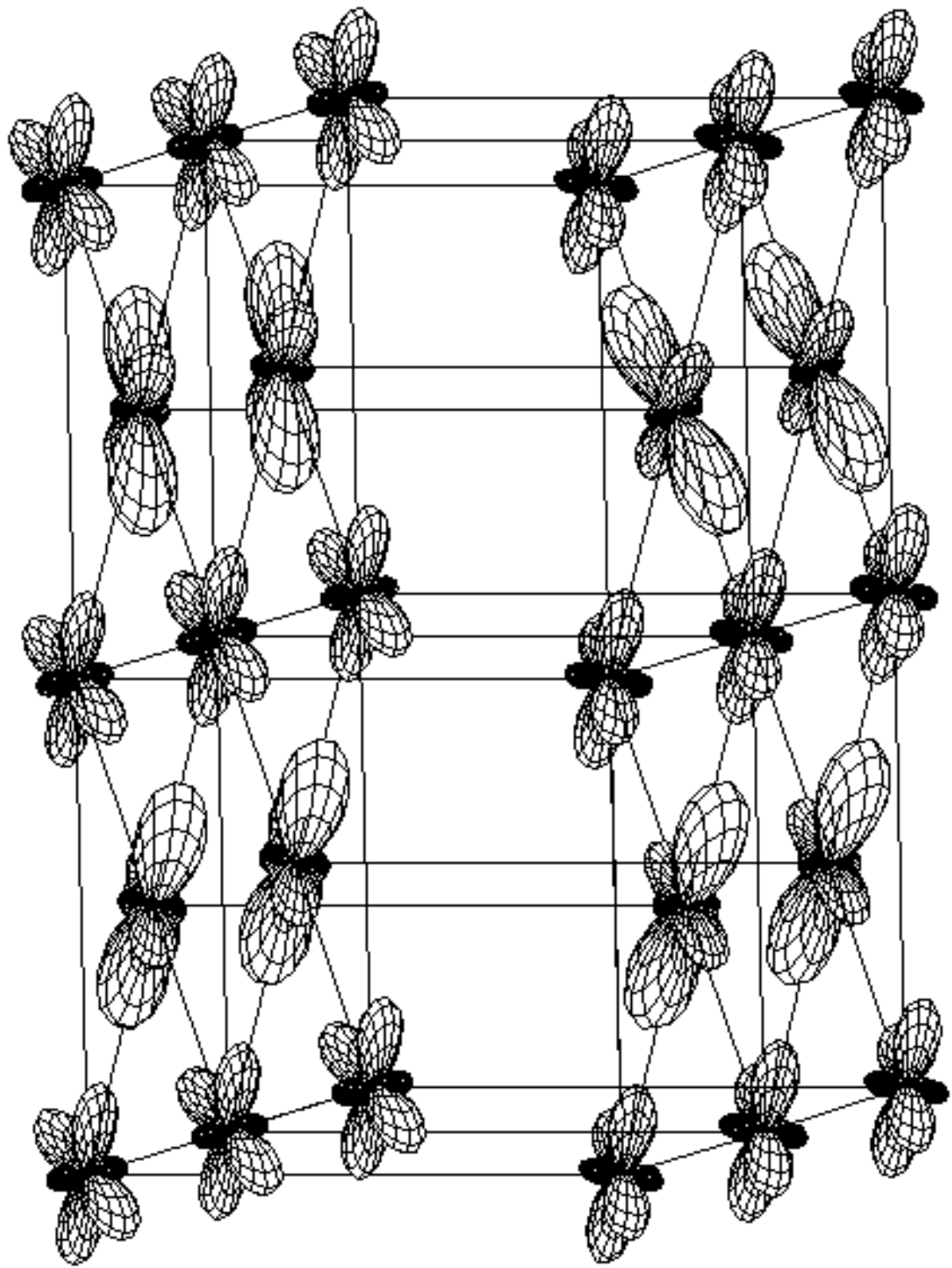
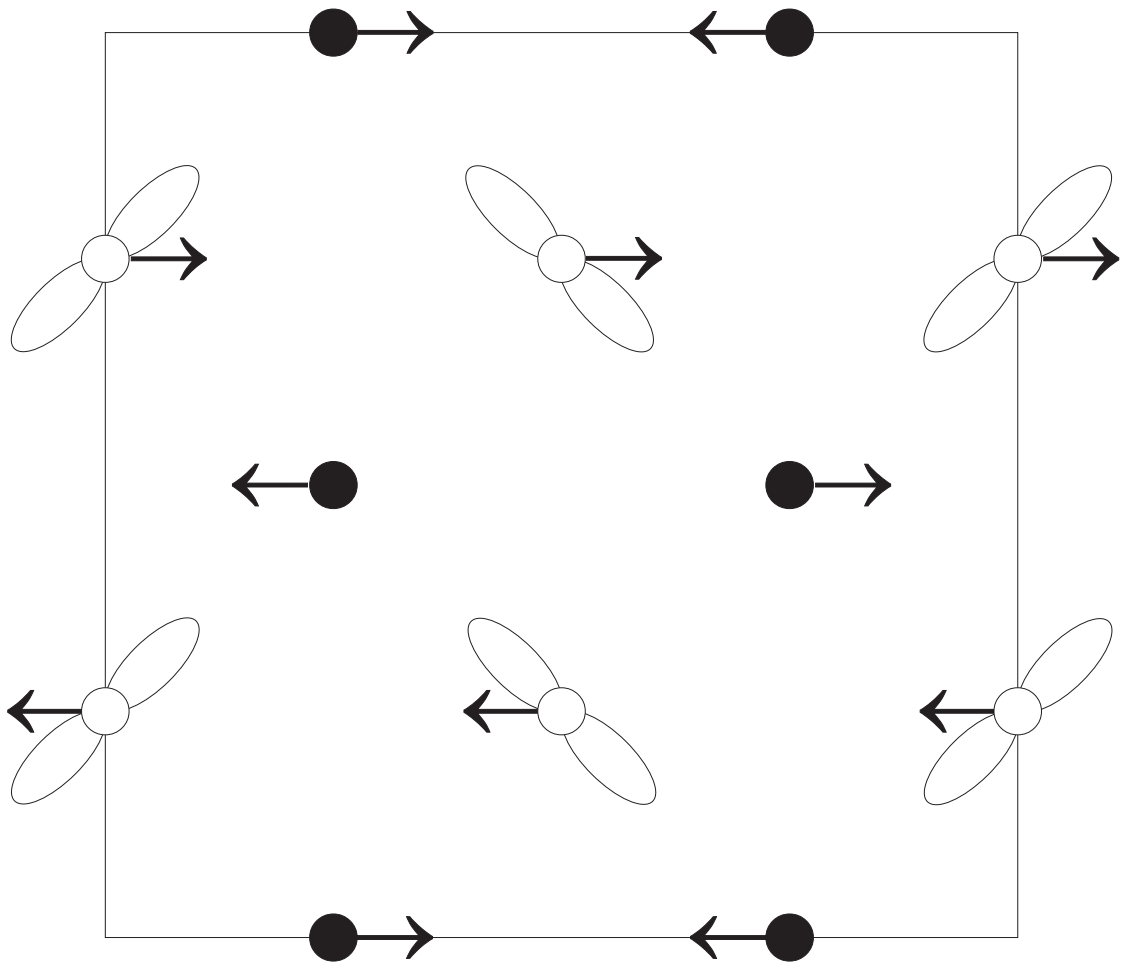
\end{document}